# Signatures for half-metallicity and nontrivial surface states in a Kagome-lattice magnetic Weyl semimetal $Co_3Sn_2S_2$

Lin Jiao[1,†,‡], Qiunan Xu[1,†], Yeryun Cheon[1], Yan Sun[1], Claudia Felser[1], Enke Liu[2,3*], Steffen Wirth[1,*]

[1]*Max-Planck-Institute for Chemical Physics of Solids, 01187 Dresden, Germany*

[2]*Institute of Physics, Chinese Academy of Sciences, Beijing 100190, China*

[3] *Songshan Lake Materials Laboratory, Dongguan, Guangdong 523808, China*

## Abstract

Weyl semimetals with time reversal symmetry breaking are expected to show various fascinating physical behaviors, such as intrinsic giant anomalous Hall effect, chiral anomaly effect in the bulks, and Fermi arcs on the surfaces. Here we report a scanning tunneling microscopy study on the magnetic Weyl semimetal candidate $Co_3Sn_2S_2$. According to the morphology and local density of states of the surface, we provide assignments to different surface terminations. The measured local density of states reveals a semimetal gap of ~300 mV, which is further verified as the gap in spin-minority bands using spin-resolved tunneling spectra. Additionally, signature for the nontrivial surface states around 50 mV is proposed. This is further confirmed by the observations of standing waves around a step-edge of the sample. Our observations and their comparison with band structure calculations provide direct yet timely evidence for the bulk and surface band structures of the magnetic Weyl semimetal $Co_3Sn_2S_2$.

[†]These authors contributed equally
[‡] Present address: Department of Physics and Frederick Seitz Materials Research Laboratory, University of Illinois Urbana-Champaign, Urbana, Illinois 61801, USA
[*]Corresponding author. Email: ekliu@iphy.ac.cn; Wirth@cpfs.mpg.de

# I. INTRODUCTION

Topological electronic states originating from a nontrivial bulk band structure have motivated immense interests in the community of condensed matter physics [1, 2]. Recently, more exotic Weyl semimetals have been proposed and discovered as a new type of topological materials [3-8]. Weyl nodes stem from the band splitting driven by spin-orbital coupling (SOC) due to the inversion symmetry breaking or time reversal symmetry (TRS) breaking. In a Weyl semimetal, the net Berry flux arises between a pair of Weyl nodes of opposite chirality, and non-closed Fermi arcs are expected to connect the projection of two opposite Weyl nodes [8]. On the contrary, the surface states in other topological materials like Dirac semimetals are closed. As a result, a Fermi arc is regarded as a fingerprint of Weyl semimetal, which can be directly verified by surface sensitive methods [6-8], such as scanning tunneling microscopy (STM) and angle-resolved photoemission spectroscopy. Although Fermi arcs have been observed in many non-magnetic Weyl semimetals [8], it is still unexamined in magnetic Weyl semimetals where breaking of TRS plays a role. In such systems, two-fold degenerated Weyl fermions are expected to generate some exotic spin-electronic phenomena, such as an anomalous Hall effect [9, 10], chiral anomaly in transport measurements [11–13], and a gravitational anomaly effect [14]. Consequently, applications of topological and spintronic properties in a magnetic Weyl semimetal are promising [15,16].

Recently, some candidate materials for TRS-breaking magnetic Weyl systems have been proposed, such as $Y_2Ir_2O_7$ [3], $HgCr_2Se_4$ [5], GdPtBi [9], YbPtBi [17], $Co_2ZrSn$ [18], $Co_2MnGa$ [19], and $Co_3Sn_2S_2$ [20–22]. Among them, semimetallic $Co_3Sn_2S_2$ is unique since an intrinsic and a large anomalous Hall effect was clearly observed by electric transport measurements [20,22]. This compound possesses a type-$I_A$ half-metallic ferromagnetism with the minority-spin component gapped around the Fermi level ($E_F$). Further density functional theory (DFT) calculations indicate that there are six nodal rings in the majority-spin channel based on the band inversion. Upon including SOC, the nodal rings are gapped and three pairs of Weyl nodes in the first Brillouin zone are formed [20-22]. More interestingly, the Weyl nodes are located only 50-60 meV above the charge neutrality point [20-22], generating pronounced effects in transport measurements. With giant anomalous Hall effect and chiral anomaly induced negative magnetoresistance, the Weyl fermions are expected to, appear in a magnetic Kagome lattice.

To further elucidate $Co_3Sn_2S_2$ as a magnetic Weyl semimetal, investigations of its (spin dependent) band structure and Fermi arc surface states are crucial. Here, we study $Co_3Sn_2S_2$ by utilizing low temperature scanning tunneling microscopy/spectroscopy (STM/STS) down to 2 K. Normal tungsten tips and spin-polarized Cr-coated STM tips are used for measuring the local density of states (LDOS) and spin-resolved LDOS, respectively. At the same time, surface states induce standing waves near a step-edge, and their dispersion is visualized by measuring STS maps around step-edges.

## II. EXPERIMENTAL METHODS

The single crystals of $Co_3Sn_2S_2$ can be grown by slowly cooling the melts with the congruent composition. The polycrystalline samples of $Co_3Sn_2S_2$ were sealed in a quartz tube with some iodine. The samples were heated to 1000 °C over 6 hours and kept there for 24 hours before being slowly cooled to 600 °C over 7 days. Thin single-crystalline flakes were obtained on the wall of the quartz tubes, and large single-crystalline plates were obtained at the bottom of the quartz tubes. $Co_3Sn_2S_2$ single crystals were cleaved *in situ* at $T < 20$ K to expose a (001) surface. After cleaving, the samples are quickly transferred to the STM head and measured in ultra-high vacuum ($p < 3 \times 10^{-9}$ Pa) and at low temperature ($T = 2$ K). The tunneling spectra were measured using tungsten tips (for normal LDOS) or Cr-coated tips (for spin-resolved LDOS) and acquired by the standard lock-in technique.

## III. RESULTS AND DISCUSSION

A. d$I$/d$V$-spectra of different crystal surfaces

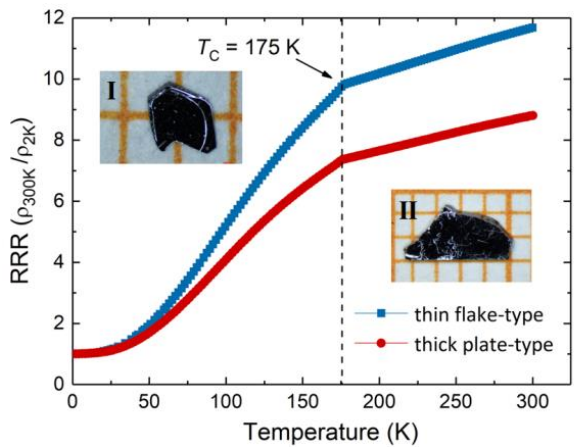

Fig. 1. Residual resistance ratio (RRR ≡ $\rho_{300K}/\rho_{2K}$) of $Co_3Sn_2S_2$ measured in zero magnetic field. Red and blue curves are measurements of thick plate-type sample and thin flake-type sample respectively. RRRs are close to 9 and 12 for each type. Insets are optical photographs of the single crystalline plate (Inset I) and thin-flake (Inset II) $Co_3Sn_2S_2$. Samples are synthesized by the means described in the methods part.

As shown in Fig. 1, $Co_3Sn_2S_2$ orders ferromagnetically with a Curie temperature ($T_C$) of ~175 K. Plate-type samples have typical residual resistance ratio (RRR) of around 9, while thin-flakes normally shows a higher RRR of around 12. Several single crystals of these two types were studied by STM at 2 K which is well inside the magnetically ordered phase. Figure 2a shows the hexagonal unit cell of $Co_3Sn_2S_2$. As demonstrated in the inset, both Sn and S-layers exhibit a hexagonal lattice with Sn-Sn and S-S distances of 5.369 Å, while Co atoms are arranged in atwo-dimensional (2D) Kagome lattice. S-, Sn-, and $Co_3Sn$-layers sandwich along c-axis and give a lattice parameter $c = 13.176$ Å.

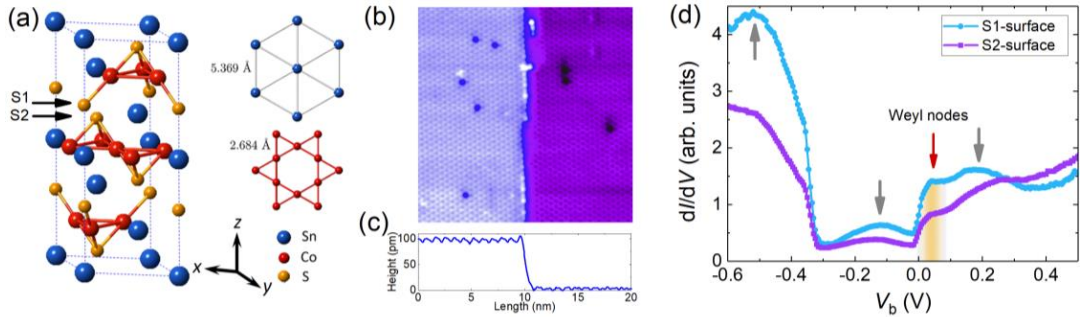

Fig. 2. (a) Unit cell of $Co_3Sn_2S_2$ in hexagonal setting. The small panels show the lattice of Sn and Co viewed along the *c*-axis. It is noted that S1, S2 and Sn-layers share the same lattice structure. (b) A $20 \times 20$ nm$^2$ topography obtained from the plate-type sample. (c) Height profile of one line cut in (b) from left- to right-hand side, indicating that the step-edge is around 90 pm in height. (d) Typical STS spectra measured on the left side (S1-surface) and right side (S2-surface) of the step-edge in (b). Gray arrows mark the peaks in the light blue curve, which locate at the same energies as those in the calculations (see Fig. 3). The filled yellow region around 50 mV above the Fermi level and the red arrow indicates energy levels where nontrivial surface states and Weyl nodes are theoretically predicted. Tunneling conditions used for all the STS measurements are $T = 2$ K, $V_b = 0.6$ V and $I_{set} = 0.5$ nA. A modulation voltage $V_{mod} = 3$ mV was applied to the sample for STS measurements.

Figures 2b and 2c show typical topographies of $Co_3Sn_2S_2$ for the thick plate-type sample and thin flake-type sample, respectively. Plate-type samples normally show only few vacancy-type defects on the surface as shown in Fig. 2b. Given that the step-edge in Fig. 2b is around 90 pm in height and the exposed lattice structure is hexagonal, the observed surfaces in this figure should be the S1- or S2-terminated surface, as denoted in Fig. 2a. To elucidate the surface termination, we measured LDOS by normal tungsten STM tip and compared the spectra with calculated density of states (DOS) for S-terminated surfaces (see Fig. S1 in Supplementary Material [23]). Figure 2d shows similar tunneling spectra obtained on each side of the step-edge in Fig. 2b. The dominant feature of the spectra is the pronounced but partially opened band gap, with a size of over 300 meV. In addition, there are several broad peaks distributed in the curves. Remarkably, the gap and peaks (marked by gray arrows in Fig. 2(d)) in the d$I$/d$V$-curves are all reproducible in the projected DOS for S-terminated surface (see Fig. S1 in Supplementary Material [23]).

Meanwhile, the calculated bulk DOS also resembles the d$I$/d$V$-curves in Fig. 2d. More importantly, while the projected DOS from our slab calculations of the first three layers as well as the bulk DOS are relatively featureless at around 50 meV, our measurements reveal an additional hump in LDOS centered around 50 mV (Fig. 2d). This energy level is where Fermi arc surface states are expected [20-22]. For the STM tip, tunneling into the 2D surface states could be much more pronounced than into the bulk states [24], while in the calculations, the contributions of the surface states are averaged out and therefore is not pronounced. Additionally, this hump is also sensitive to defect, distinguishing from the bulk states (see Fig. S2 in Supplementary Material [23]). Consequently, we temporally propose that the Fermi arcs induce the additional hump around 50 mV in the d$I$/d$V$-curves. This argument will be further elucidated in the following.

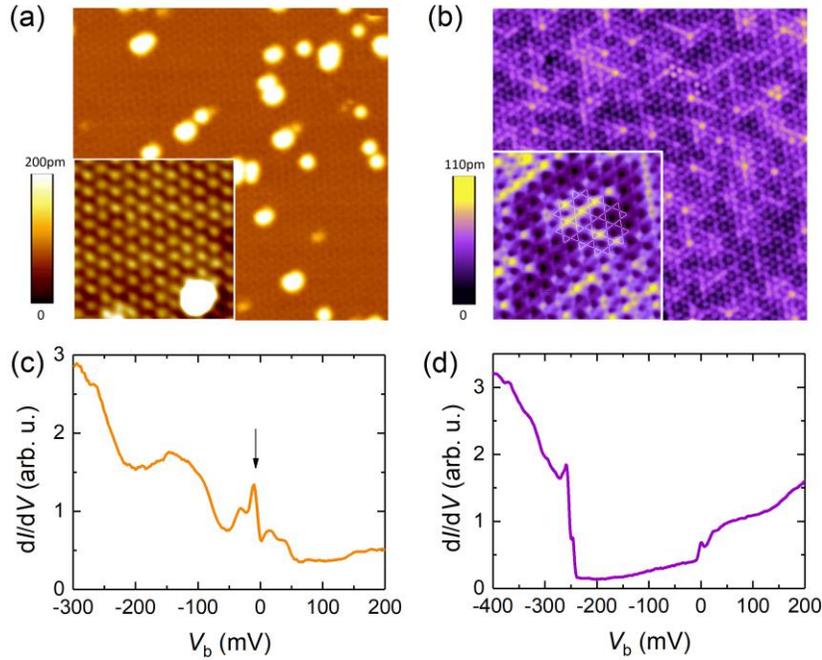

Fig. 3 (a) and (b) are 20 × 20 nm$^2$ Sn-terminated surface and the Kagome lattice of Co, respectively. Magnified 5 × 5 nm$^2$ topographies of the corresponding surface are attached to the figures as insets. Sn-terminated surface shows triangular lattice while Co arrays in a Kagome lattice as indicated by the white net. (c) and (d) present typical d$I$/d$V$-curves obtained on (a) and (b), respectively. Data was collected around 2 K.

We noted that the d$I$/d$V$-curves in Fig. 2 (d) are widely observed in the sample, while in some cases other types of surfaces and spectra are observed. As shown in Fig. 3(a), although atoms also array in a triangular lattice, the types of defect are different from the S-terminated surfaces. In the latter case, a high density of vacancy is observed, while only big protrusions present in Fig.

3(a). Correspondingly, the d$I$/d$V$-spectra are quite different from S-terminated surfaces, as shown in Fig. 3(c). Taking the lattice structure and the spectrum into account, we define such rarely observed surface in Fig. 3(a) to be Sn-terminated surface. This definition is identical to Morali *et al*. [25]. The sharp peak marked by the black arrow in Fig. 3(c) seems identical to that observed by Ying *et al*. [26], which is understood as a feature of frustrated Kagome flat band. But this kind of surface is defined as S-terminated surface in Ref. [26]. The flake-type crystal was also investigated. This sample normally cleaves between $Co_3Sn$-S layers, and therefore, the Kagome lattice is exposed (see Fig. 3(b)). The d$I$/d$V$-spectrum in Fig. 3(d) is obtained on the Kagome lattice. Surprisingly, this spectrum shares the major features with those from S-terminated surface, except that the partially opened bulk gap reduces from 300 to 250 meV. Additionally, small peaks around $E_F$ and -250 mV are observed, which are likely origin from the 3$d$ orbitals of Co. However, Sn is normally missing in the exposed $Co_3Sn$ layer and the Kagome lattice becomes inhomogeneous electronically, as shown in Fig. 3(b). Therefore, the following STS measurements are exclusively focused on the surfaces of thick plate samples.

## B. Half metallicity

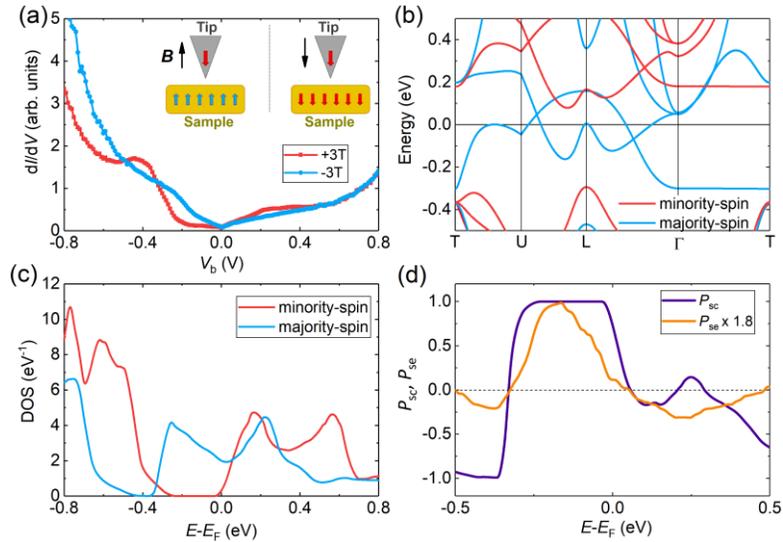

Fig. 4. (a) Spin resolved d$I$/d$V$-curves of $Co_3Sn_2S_2$ measured at magnetic fields of +3 T and −3 T using an antiferromagnetic Cr-coated tip. The magnetic field is along the *c*-axis of the sample. The inset shows the configuration of the spins of the electrons in the tip apex and majority-spin bands of the sample versus field orientations. The spin orientation in the antiferromagnetic tip does not change with external magnetic field. Tunneling conditions are $V_b$ = 1000 mV, $I_{set}$ = 1.0 nA and $V_{mod}$ = 3 mV. The spectra were collected around 2 K. (b) and (c) The energy band and DOS calculations without SOC for bulk. Blue and red curves represent majority-spin and minority-spin channels, respectively. (d) Comparison of the spin polarization ratio of the tip-sample junction between experimental ($P_{se}$) and calculated ($P_{sc}$) results.

One key feature of $Co_3Sn_2S_2$ is its half-metallic nature, with one spin channel being gapped and the other being metallic [20, 27-30]. This property is of great importance for the magnetic Weyl semimetal with observable topological bands and measurable electronic transport properties such as the giant anomalous Hall or Nernst effect. Here, we further investigate its spin-dependent band structure by means of spin-resolved STS using a Cr-tip [31,32]. This tip uses a thin layer of Cr (∼10 nm) to provide predominantly out-of-plane spin resolution. Such an antiferromagnetic Cr-coating layer has additional advantages, compared to ferromagnetic tips: it produces no significant stray field and is insensitive to external fields. Detailed information and the preparation process of this spin-polarized tip can be found in Ref. [33]. Figure 4(a) shows the spin-resolved STS measured around 2 K and with ± 3 T magnetic field applied along the *c*-axis of the sample. The data taken at +3 T after the cycle is consistent with the initial measurement at +3 T (see Fig. S3 in the Supplementary Material [23]), indicating the tip status is the same during the measurements. For the antiferromagnetic Cr-tip, we assume its spin at the tip apex points downward. The spin orientation of the tip should not change with external field orientation, because of its antiferromagnetic order. Therefore, one of the two spin-degenerated bands dominate the d$I$/d$V$-curves either at +3 T or −3 T due to the spin-valve effect [34]. Comparing with the experimental data, the calculations in Figs. 4(b) and (c) reproduce the minority-spin channel very well. Especially around $E_F$, the ~300 meV spin gap is well confirmed. On the other hand, the majority-spin bands can also be detected for the negative field direction, which is manifested by the 'V'-shape DOS around $E_F$. However, the spin gap of the majority-spin band around -0.5 eV could not be clearly resolved, and only a flat region is presented in the d$I$/d$V$-curve. Because the spin polarization of the Cr-tip, $S$, is not 100% (typically around∼50% [31]) and reaches its maximum around $E_F$, one should get a better spin resolution at low bias voltage. We also note that spins of the tip and the sample are often not perfectly aligned with each other.

With the spin dependent d$I$/d$V$-curves, one can further calculate the effective spin polarization ratio ($P_{se}$) of the tip-sample tunnel junction [34]:

$$P_{se} = \frac{1}{S}(dI/dV_{\uparrow\uparrow} - dI/dV_{\uparrow\downarrow})/(dI/dV_{\uparrow\uparrow} + dI/dV_{\uparrow\downarrow}),$$

where d$I$/d$V_{\uparrow\uparrow}$ and d$I$/d$V_{\uparrow\downarrow}$ are the tunneling conductance of the majority- and minority-spin components, respectively, and $S$ is the spin polarization ratio of the tip. Theoretically, the spin polarization of the sample can be derived by $P_{sc}$= (DOS↑-DOS↓)/(DOS↑ +DOS↓), where ↑ and ↓ denote the majority- and minority-spin components. Taking $S$ = 55%, $P_{se}$ fits $P_{sc}$ in Fig. 4(d).

Remarkably, the experimental data follows nicely the calculations below $E_F$ and the value of $S$ is comparable with previous report [35]. Above $E_F$, where the experimental d$I$/d$V$-curves are relatively featureless, we notice that $P_{se}$ and $P_{sc}$ have opposite peak and valley positions, such bias dependent tunneling probability can be understood in terms of bias-dependent spin polarization of the tip [36]. The field-dependent tunneling and the similarity between $P_{se}$ and $P_{sc}$ in a wide energy range confirms the half-metallic nature of $Co_3Sn_2S_2$, though the current method could not derive a precisely spin polarization ratio around the Fermi level.

C. Analysis of the surface states

The above experimental results confirm that our bulk band structure calculations describe the bulk states of $Co_3Sn_2S_2$ appropriately, even after including the spin components. Now, we turn to the analysis of the surface states. In this respect, STM can be applied to probe the band dispersion by making use of point- (e.g. an impurity or vacancy) or line-defects (e.g. step-edges). In these cases, standing waves of the electrons due to the so-called Friedel oscillations are expected [37, 38]. Normally such oscillations are pronounced for 2D states in which only in-plane scatterings are allowed [39]. Figure 5(a) shows a topography of $Co_3Sn_2S_2$, likely of a S-terminated surface, with several line-defects along the $x$-direction. A height scan along the white dashed-dotted line is presented in Fig. S4 in the Supplemental Material [23], which shows the step-edge on the right-hand side is a half unit cell (7 Å) in height. On the left-hand side, several scratches due to missing atoms on the top layer could be observed. To probe standing waves of the surface states, we measured the LDOS along the white dash-dotted line and between the two edges.

The raw data of line-scanned LDOS are included in Fig. S5 in the Supplementary Material [23], while Fig. 5(b) shows the residual LDOS $\Delta \frac{dI}{dV(r,V)}$ with the lattice-induced periodic oscillations and an averaged background signal filtered out. Here we define $\Delta \frac{dI}{dV(r,V)} \equiv H_L[\frac{dI}{dV(r,V)}, \omega] - \frac{1}{n}\sum_{r=0}^{15} H_L[\frac{dI}{dV(r,V)}, \omega]$, where $r$ is the distance to the deep step-edge, $n$ is the total number of the averaged curves and $H_L[f(x), \omega]$ is a low-pass filter with a cut-off frequency $\omega$. In Fig. 5(b) one can find that the residual LDOS is featureless below $E_F$, except a peak around the step-edges. Starting from $V_b = 10$ mV, standing waves appear and extend up to around 80 mV. This energy region is where topological surface states are predicted based on Fig.

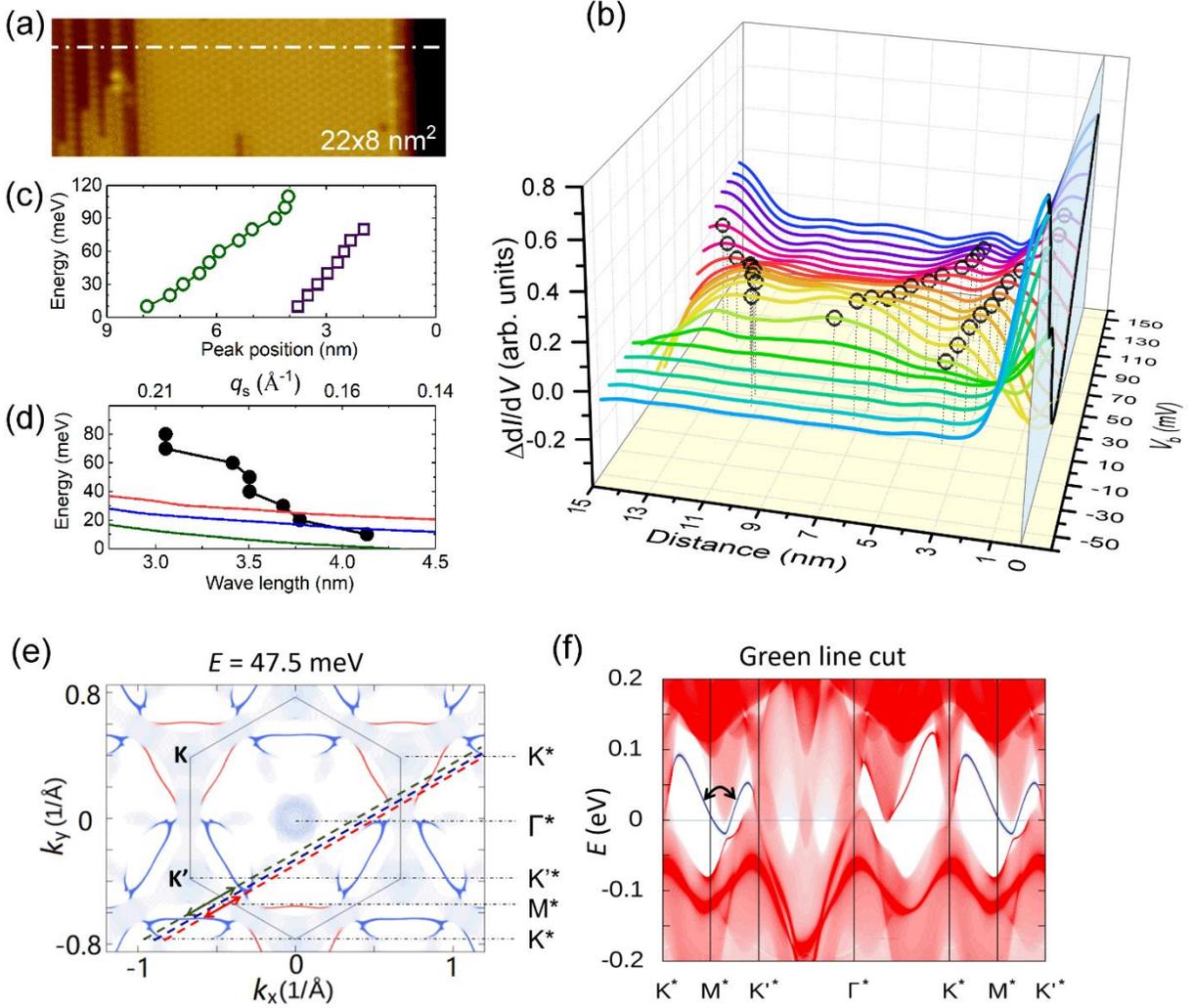

Fig. 5. (a) 22 × 8 nm² STM topography of $Co_3Sn_2S_2$ with several step-edges. (b) Residual LDOS (with the lattice-induced waves and an averaged background filtered out) at selected bias voltages along y-direction (the white dash-dotted line) in (a). The black open circles mark the peak position of the standing waves. A light blue cross section in the d$I$/d$V$-$V_b$ plane at $x = 0$ nm shows the LDOS at the step-edge by the black curve. (c) Peak positions of the residual LDOS (black circles in (c)) versus bias voltage. (d) Derived wavelength and scattering wavevectors of the standing waves from (c). The solid lines are possible scattering wavelength between the nontrivial surface states derived from different cuts of the band structure as marked in (e). (e) Calculated DOS at 47.5 meV with only the surface states highlighted. Red and blue bands are trivial and nontrivial surface states, respectively. The three dashed lines indicate three cuts parallel to ΓK-direction, which is perpendicular to the step-edges. Scatterings between the adjacent nontrivial surface states are marked by the solid arrows. We note that (d) and (e) share the same color code. The definition of the high symmetry points (Γ, M, K) and their derived positions (Γ*, M*, K* are the corresponding intersections of the horizontal gray dashed lines and the other dashed lines) in different cuts are also labeled in (e). (f) Calculated band structure of $Co_3Sn_2S_2$ along the green dashed line cut in e. The blue curves are nontrivial surface state bands. The black arrow is putative scattering wavevector between the nontrivial surface states parallel to Γ*K* direction.

2. Tracing the peak positions of the standing waves indicates an energy dispersion of the standing waves, as shown in Figs. 5(b) and (c). Based on the wavelength ($l_s$), in Fig. 5(d) we calculate the scattering wave vector $q_s = 2\pi/l_s$. As the standing waves are pronounced around the Weyl nodes, it is natural to compare $q_s$ with the potential scattering path between the adjacent surface states. The potential scattering path is derived based on the band structure (along certain cut) and Fermi surface as plotted in Figs. 5(e) and 5(f). The band structure conveys some important messages: Firstly, the nontrivial surface states (in blue color) are dominant around 50 (±50) meV as expected; Secondly, the parabolic-like surface-state-bands are prone to host scattering around the step-edges. Thirdly, part of the trivial surface states (red curves in (f) from Γ* to K*) merge into the bulk states, which is not likely to give clear Friedel oscillation around the step-edge. Therefore, the observed Friedel oscillation is understood as the scattering between the nontrivial surface states. In Fig. 5(d) the measured dispersion of $q_s$ (black dots) are compared to the calculated ones between two nontrivial surface states (indicated by the black arrow in Fig. 5(f)). Although putative scattering paths in calculations and the STS data indeed share similar behavior, there is considerable deviation between them. However, taking the complicated FS of this compound and the small energy window into account, it is difficult to precisely determine the details of the surface states in our model. A much intensive calculation is required for further study once additional information from momentum space is available.

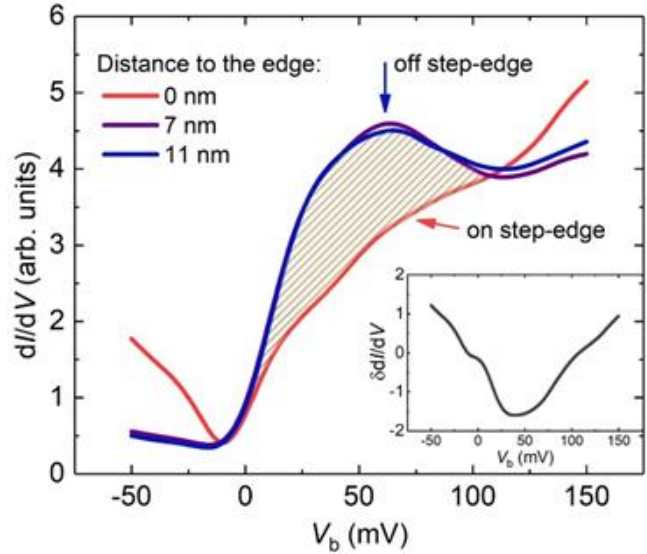

Fig. 6. d$I$/d$V$-curves on top of and away from the deep step-edge in Fig. 5(a). The red curve is obtained right on top of the step-edge. The inset shows the difference of LDOS between the red and blue curves ($\delta$d$I$/d$V$= d$I$/d$V(r=0)$- d$I$/d$V(r=7)$).

## IV. CONCLUSION AND PERSPECTIVES

Taking the above experimental results and calculations, we provide a new insight into the surface and bulk states of $Co_3Sn_2S_2$. Firstly, the surface terminations and the corresponding tunneling spectra have been verified. Then, the comparison between STS spectrum and band structure calculations provides signatures of Weyl nodes in this ferromagnetic compound. Particularly, the step-edge shows an important role for extending the scope of topological nontrivial states. In this respect, STM is powerful for studying diverse topological nontrivial states in low dimension or at the edge, corner or hinge. In respect to material, it is worth mentioning that the 2D Kagome lattice supports exotic phases including spin liquid [40], topological insulator [41], and quantum anomalous Hall effect [42]. While the Kagome-lattice metallic $Fe_3Sn_2$ was recently found to bear massive Dirac fermions [43], the semimetallic $Co_3Sn_2S_2$ now is promising to host Weyl fermions, which further manifests the Kagome lattice as a versatile platform for novel electronic states, such as a Kagome magnet [26]. Another intriguing feature is the suppression of LDOS just around 50 mV on the step-edge, where nontrivial surface states are predicted. After removing the signal obtained away from the edge, a gap-like feature is manifested in the δd$I$/d$V$-curves right on the edge (black curves in Figs. 5c and Fig. 6). Whether chiral edge states will replace the Fermi arcs on the step-edge in the case of quantum anomalous Hall insulator is still an open question and need further study [44].

## ACKNOWLEDGEMENTS

E.L. acknowledges the National Key R&D Program of China (No. 2017YFA0206303) and the National Natural Science Foundation of China (No. 51722106). Financial support from the Deutsche Forschungsgemeinschaft (DFG) within the Schwerpunktprogramm SPP1666 and ERC Advanced Grant (No. 742068) 'TOPMAT' are gratefully acknowledged.

# References

[1] M.Z. Hasan and C.L. Kane, Rev. Mod. Phys.**82**, 3045-3067 (2010).

[2] X. Qi and S. Zhang, Rev. Mod. Phys.**83**, 1057-1110 (2011).

[3] X. Wan, A.M. Turner, A. Vishwanath, and S.Y. Savrasov, Phys. Rev. B**83**, 205101 (2011).

[4] L. Balents, Physics**4**, 36 (2011).

[5] G. Xu, H. Weng, Z. Wang, X. Dai, and Z. Fang, Phys. Rev. Lett.**107**, 186806 (2011).

[6] S.Y. Xu, I. Belopolski, N. Alidoust, M. Neupane, G. Bian, C. Zhang, R. Sankar, G. Chang, Z. Yuan, C. Lee, S. Huang, H. Zheng, J. Ma, D. S. Sanchez, B. Wang, A. Bansil, F. Chou, P. P. Shibayev, H. Lin, S. Jia, and M. Z. Hasan,Science**349**, 613-617 (2015).

[7] B.Q.Lv, H. M. Weng, B. B. Fu, X. P. Wang, H. Miao, J. Ma, P. Richard, X. C. Huang, L. X. Zhao, G. F. Chen, Z. Fang, X. Dai, T. Qian, and H. Ding,Phys. Rev. X**5**, 031013 (2015).

[8] See e.g. M.Z. Hasan, S. Xu, I. Belopolski, and S. Huang, Annu. Rev. Condens. Matter Phys.**8**, 289-309 (2017) and reference therein.

[9] T.Suzuki, T. Suzuki, R. Chisnell, A. Devarakonda, Y.-T. Liu, W. Feng, D. Xiao, J. W. Lynn & J. G. Checkelsky,Nat. Phys.**12**, 1119 (2016).

[10] S. Nakatsuji, N. Kiyohara, and T. Higo, Nature**527**, 212 (2015).

[11] H. Nielsen and M. Ninomiya, Phys. Lett. B**130**, 389 (1983).

[12] D.T. Son and B.Z. Spivak, Phys. Rev. B**88**, 104412 (2013).

[13] X. Huang, L. Zhao, Y. Long, P. Wang, D. Chen, Z. Yang, H. Liang, M.Xue, H. Weng, Z. Fang, X. Dai, and G. Chen, Phys. Rev. X **5**, 31023 (2015).

[14] J.Gooth, A. C. Niemann, T. Meng, A. G. Grushin, K. Landsteiner, B.Gotsmann, F. Menges, M. Schmidt, C.Shekhar, V. Süß, R.Hühne, B.Rellinghaus, C. Felser, B. Yan, and K.Nielsch,Nature**547**, 324 (2017).

[15] C.-K.Chan, P. A.Lee, K. S. Burch, J. H.Han, and Y. Ran, Phys. Rev. Lett.**116**, 026805 (2016).

[16] Y. Tokura, M. Kawasaki, and N. Nagaosa, Nat. Phys. **13**, 1056–1068 (2017).

[17] C. Y. Guo, F. Wu, Z. Z. Wu, M. Smidman, C. Cao, A. Bostwick, C. Jozwiak, E. Rotenberg, Y. Liu, F. Steglich & H. Q. Yuan,Nat. Commun. **9**, 4622 (2018).

[18] Z. Wang, M. G. Vergniory, S. Kushwaha, M. Hirschberger, E. V. Chulkov, A. Ernst, N. P. Ong, Robert,J. Cava, and B. A.Bernevig,Phys. Rev. Lett. **117**, 236401 (2016).

[19] A. Sakai, Y. P.Mizuta, A. A. Nugroho, R.Sihombing, T.Koretsune, M. -T. Suzuki, N.Takemori, R. Ishii, D.Nishio-Hamane, R.Arita, P.Goswami, and S.Nakatsuji,Nat. Phys.**14**, 1119-1124 (2018).

[20] E. Liu, Y. Sun, N. Kumar, L.Muechler, A. Sun, L. Jiao, S. -Y. Yang, D. Liu, A. Liang, Q. Xu, J.Kroder, V. Süß, H. Borrmann, C.Shekhar, Z. Wang, C. Xi, W. Wang, W. Schnelle, S. Wirth, Y. Chen, S. T. B. Goennenwein, and C. Felser,Nat. Phys.**14**, 1125-1131 (2018).

[21] Q. Xu, E. Liu, W. Shi, L.Muechler, J. Gayles, C. Felser, and Y. Sun, Phys. Rev. B**97**, 235416 (2018).

[22] Q. Wang, Y. Xu, R. Lou, Z. Liu, M. Li, Y. Huang, D. Shen, H. Weng, S. Wang, and H. Lei,Nat. Commun. **9**, 3681 (2018).

[23] See Supplemental Material for detailed description of the methods and the data.

[24] J. Fernández, M. Moro-Lagares, D. Serrate, and A. A. Aligia, Phys. Rev. B **94**, 75408 (2016).

[25] N. Morali, R. Batabyal, P. K. Nag, E.K. Liu, Q. N. Xu, Y. Sun, B. H. Yan, C. Felser, N. Avraham, H. Beidenkopf, arXiv: 1903.00509

[26] J.X. Yin, S. S. Zhang, G. Q. Chang, Q. Wang, S. S. Tsirkin, Z. Guguchia, B. Lian, H. B. Zhou, K. Jiang, I. Belopolski, N. Shumiya, D. Multer, M. Litskevich, T. A. Cochran, H. Lin, Z. Q. Wang, T. Neupert, S. Jia, H. C. Lei, and M. Z.Hasan, Nat. Phys. (2019) https://www.nature.com/articles/s41567-019-0426-7

[27] R.Weihrich, I.Anusca, and M. Zabel, Z. Anorg. Allg. Chem. **631**, 1463–1470 (2005).

[28] W. Schnelle, A. Leithe-Jasper, H. Rosner, F. M. Schappacher, R. Pöttgen, F. Pielnhofer, and R. Weihrich, Phys. Rev. B **88**, 144404 (2013).

[29] Y. S. Dedkov, M. Holder, S. L. Molodtsov, and H. Rosner, J. Phys. Conf. Ser. **100**, 072011 (2008).

[30] M. Holder, Yu. S. Dedkov, A. Kade, H. Rosner, W. Schnelle, A. Leithe-Jasper, R. Weihrich, and S. L. Molodtsov, Phys. Rev. B **79**, 205116 (2009).

[31] R.Wiesendanger, Rev. Mod. Phys.**81**, 1495-1550 (2009).

[32] H. Oka, O. O. Brovko, M.Corbetta, V. S. Stepanyuk, D. Sander, and J. Kirschner,Rev. Mod. Phys.**86**, 1127 (2014).

[33] Detailed information of the Cr-coated tip is available from NaugaNeedles LLC: http://nauganeedles.com/products-USSTM-W500-Cr

[34] M.Bode, M.Getzlaff, andR.Wiesendanger, Phys. Rev. Lett. **81**, 4256-4259 (1998).

[35] M.Corbetta, S.Ouazi, J.Borme, Y. Nahas, F. Donati, H. Oka, S. Wedekind, D. Sander, and J. Kirschner, Jpn. J. Appl. Phys.**51**, 30208 (2012).

[36] R. Wiesendanger, D. Bürgler, G. Tarrach, A. Wadas, D. Brodbeck, and H.−J. Güntherodt,J. Vac. Sci. Technol. B **9**, 519 (1991).

[37] J.Friedel, Philos. Mag. **43**, 153-189 (1952).

[38] M.F.Crommie, C.P.Lutz, andD.M.Eigler, Nature **363**, 524 (1993).

[39] L.Jiao, Q.N.Xu,Y.P.Qi,S.C.Wu, Y.Sun, C.Felser,and S.Wirth,Phys. Rev. B**97**, 195137 (2018).

[40] M.X. Fu, T. Imai, T.-H. Han, and Y.S. Lee, Science**350**, 6261 (2015).

[41] H.-M. Guo and M. Franz, Phys. Rev. B**80**, 113102 (2009).

[42] G. Xu, B. Lian, and S.C. Zhang, Phys. Rev. Lett. **115**, 186802 (2015).

[43] L.Ye, M. Kang, J. Liu, F. von Cube, C. R. Wicker, T. Suzuki, C.Jozwiak, A.Bostwick, E. Rotenberg, D. C. Bell, L. Fu, R.Comin, and J. G. Checkelsky, Nature **555**, 638–642 (2018).

[44] L. Muechler, E. K. Liu, Q. Xu, C. Felser, and Y. Sun, arXiv:1712.08115.